\documentclass[epj]{svjour}
\usepackage{graphicx}
\begin{document}
\title{Latest Results on the Hot-Dense Partonic Matter at RHIC}
\author{Michael J. Leitch
}                     
%
%
\institute{Los Alamos National Laboratory, Los Alamos NM 87545 USA} 
\date{Received: date / Revised version: date}
%
\abstract{
At the Relativistic Heavy Ion Collider (RHIC) collisions of heavy ions at
nucleon-nucleon energies of 200 GeV appear to have created a new form of
matter thought to be a deconfined state of the partons that ordinarily are
bound in nucleons. We discuss the evidence that a thermalized partonic
medium, usually called a Quark Gluon Plasma (QGP), has been produced. Then
we discuss the effect of this high-density medium on the production of
jets and their pair correlations. Next we look at direct photons as a
clean electro-magnetic probe to constrain the initial hard scatterings.
Finally we review the developing picture for the effect of this medium
on the production of open heavy quarks and on the screening by the QGP of
heavy-quark bound states.
} 
\maketitle

\section{Introduction}
\label{intro}
The collisions of high-energy heavy ions are thought to form very high temperatures
and densities similar to those that occurred in the earliest stages of our
universe. These collisions are thought to create matter with energy densities up to 30 times normal
nuclear matter density in a small region ($\sim10^{-14}$ m) and for a short
time ($\sim 10^{-23}$ s). Different experimental probes examine different stages of
the matter as it expands and evolves back to normal matter. Hard probes created
in the initial collisions before thermalization of the medium probe the medium through
their final-state interactions. Soft probes come from the medium itself and provide a
picture of its thermalization and spatial evolution.

Here we will review the most significant results from the RHIC heavy-ion program
and discuss their implications in terms of the nature of the hot-dense matter that
is created in these collisions. I thank many colleagues at RHIC for helping me prepare
this review, especially Akiba, Constantin, d'Enterria, Granier de Cassagnac, Jacak,
Nagle, Seto, and Zajc.  

\section{RHIC and its Detectors}
\label{sec:1}
The Relativistic Heavy Ion Collider (RHIC) collides Au ions at center-of-mass energies
per nucleon pair of up to 200 GeV, substantially higher in energy than the 17 GeV at the
CERN SPS, but lower than the 5.5 TeV anticipated at the LHC. At RHIC four experiments,
two large (PHENIX and STAR) and two small (BRAHMS and PHOBOS) experiments observe the
collisions and together have advanced our understanding considerably since the first
collisions at RHIC. Each experiment has different strengths with BRAHMS having two classic
dipole spectrometers, PHOBOS focusing on extensive silicon-strip detectors,
\break
PHENIX being a complex many-subsystem detector optimized for rare probes, and
STAR centered around a large solid angle TPC.

In heavy-ion collisions, an important aspect one studies for all observables is their dependency
on the centrality of the collision, i.e. whether a particular collision is ``head-on"
(central or e.g. 0-10\% centrality) or peripheral (e.g. 80-90\%) with only the edges
of the two colliding nuclei passing through each other. An approximate centrality measurement for
each collision is obtained with a combination of the measured charged particle multiplicity
and the yield of spectator neutrons that are observed in zero-degree calorimeters (ZDC).
These measurements are related to the actual centrality using a simple Glauber
model of the collision.

%
%

%

\section{Thermalization}
\label{sec:2}
One of the key issues in high-energy heavy-ion collisions is the time scale and
the degree to which the available energy is thermalized into a medium. Several of
the soft-sector observables give us important information on this aspect of the
collisions.

\begin{figure}
\begin{center}
  \includegraphics[height=60mm]{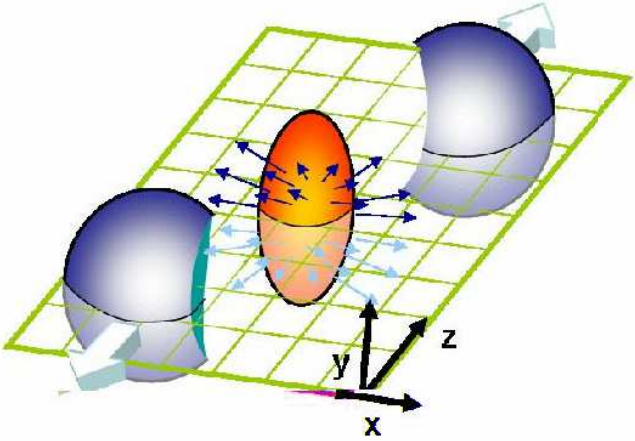}
\caption{Cartoon showing the initial geometrical asymmetry in a non-central
heavy-ion collision.}
\label{fig:flow_geometry}       
\end{center}
\end{figure}

The first is the extent to which the observed lower-momentum produced particles
exhibit ``hydrodynamic flow" as would be expected from a thermalized medium. For
not fully head-on collisions, as shown in Fig.~\ref{fig:flow_geometry}, an asymmetric
or almond shaped collision region is produced. This initial anisotropy is converted
into a corresponding momentum anisotropy which results in an azimuthal asymmetry
relative to the reaction plane in the
yield of produced particles at a given momentum. The efficiency of this conversion depends
on the properties of the medium, with large asymmetries corresponding to early
thermalization. This asymmetry is usually represented by ``$v_2$" which is the $2^{nd}$
fourier coefficient of the momentum anisotropy,

\begin{equation}
dn/d\phi \sim 1 + 2 v_2(p_T) cos(2\phi) + ...
\end{equation}

\begin{figure}
  \includegraphics[height=60mm]{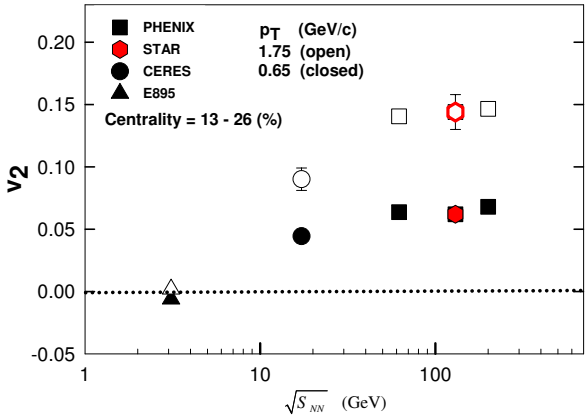}
\caption{Dependence of the flow or $v_2$ of hadrons on center-of-mass energy, $\sqrt{s_{NN}}$,
for two different transverse momentum ($p_T$) values\cite{flow_saturation}.}
\label{fig:v2_roots}       
\end{figure}

The observed flow tends to saturate near RHIC energies\cite{flow_saturation},
as shown in Fig.~\ref{fig:v2_roots}, suggesting that early thermalization has been achieved.
Also supporting early thermalization are comparisons with hydrodynamics calculations
which indicate that the flow observed at RHIC is near the maximal flow predicted by the models.
However ongoing improvements in the hydrodynamic models may shed new light on what the
true limits are.

\begin{figure}
  \includegraphics[height=60mm]{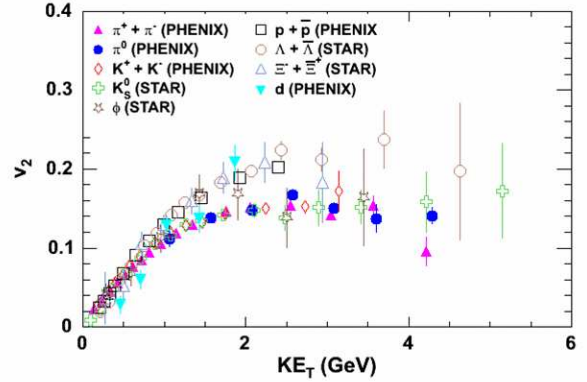}
\caption{Flow for hadrons vs. transverse energy, $KE_T = m_T - mass$.}
\label{fig:keta}       
\end{figure}

\begin{figure}
  \includegraphics[height=60mm]{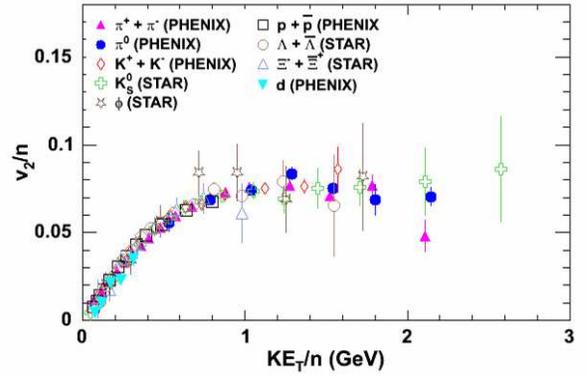}
\caption{Flow per quark vs. transverse energy per quark, $KE_T/n$.}
\label{fig:ketb}       
\end{figure}

Another important aspect of the flow measurements is that when comparisons of the flow for different
particles are made there are substantial differences observed, see Fig.~\ref{fig:keta}.
However if one plots the flow normalized to the number of valence quarks in the observed
hadron vs. the transverse kinetic energy per quark, Fig.~\ref{fig:ketb}, a universal
behavior emergences. This supports a picture where the relevant degrees of freedom of the
thermalized medium are quarks, i.e. the thermalization occurs when the matter is not hadrons
but consists of deconfined quarks. 


\begin{figure}
  \includegraphics[width=\linewidth]{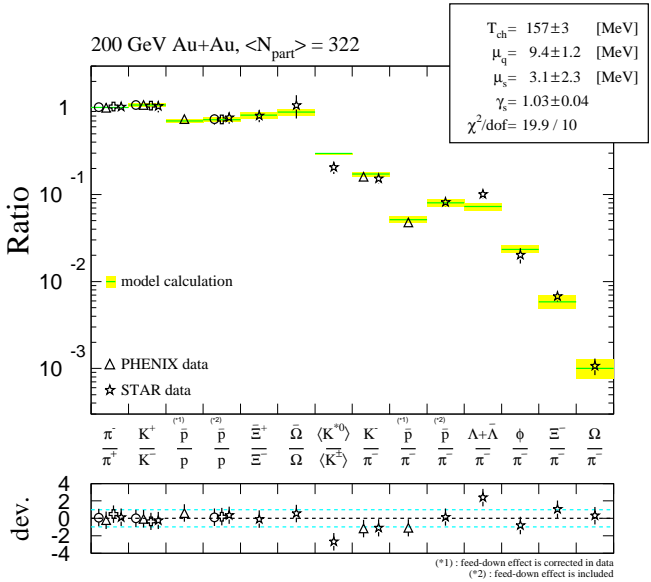}
\caption{Ratios of hadron yields are well reproduced by statistical combination models based on quark
degrees of freedom.}

\label{fig:hadronic_abundances}       
\end{figure}

Finally, statistical models that reproduce ratios of hadron yields\cite{hadron_ratios}, such
as those in Fig.~\ref{fig:hadronic_abundances}, also support a picture where the degrees of freedom
are quarks up to the ``freezeout" time when the observed hadrons form.

\section{Jets and Energy Loss}
\label{sec:3}
\begin{figure}
 \includegraphics[width=\linewidth]{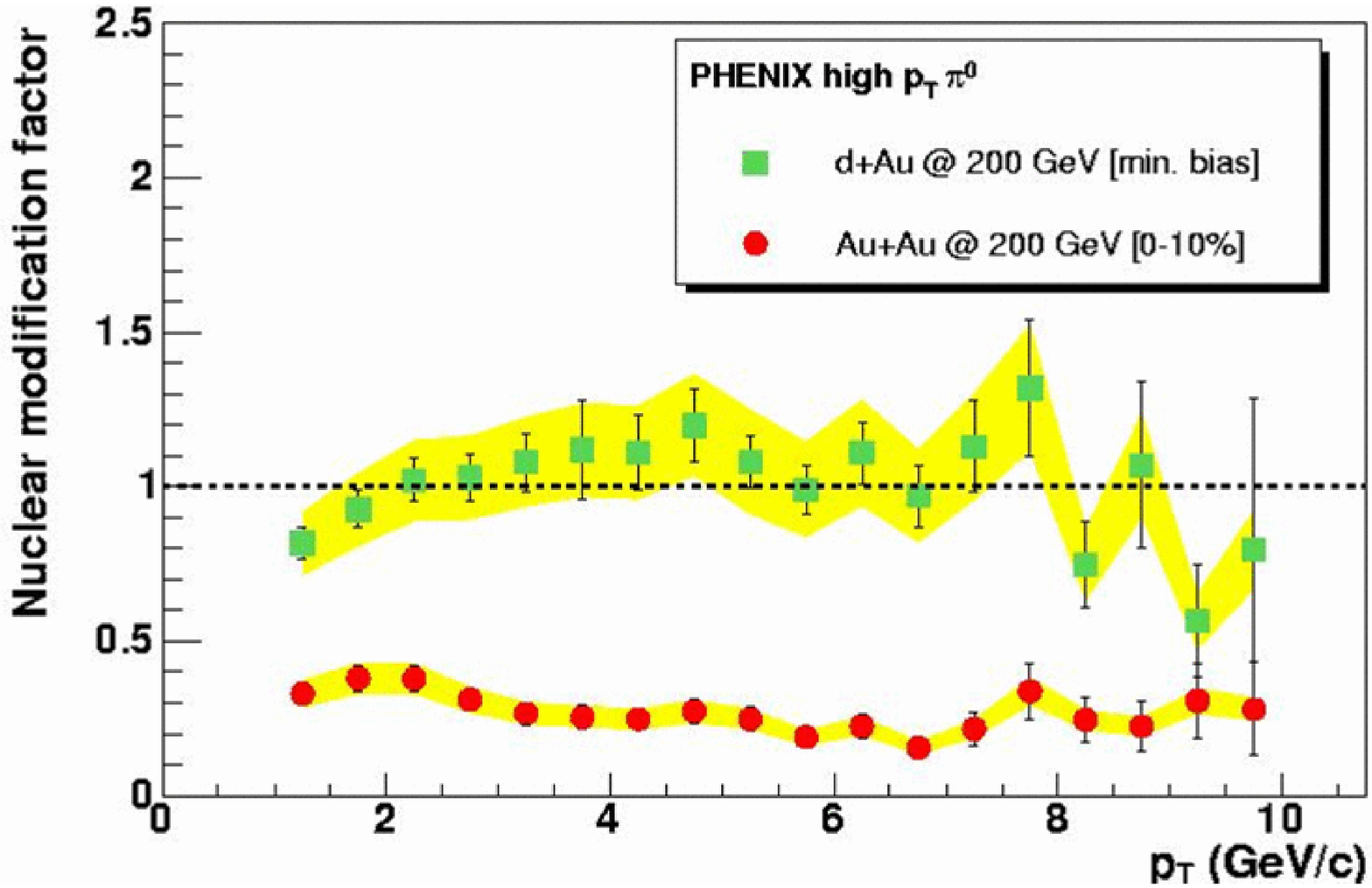}
\caption{Large suppression of the $\pi^0$ yield per binary collision in Au+Au
collisions (red circles) contrasted with lack of
suppression for d+Au collisions (green squares)\cite{phenix_jet_suppression}.}
\label{fig:jet_suppression}       
\end{figure}

Arguably the most dramatic effect seen at RHIC so far, is the strong suppression of high-$p_T$
hadrons in the most central Au+Au collisions compared to p+p collisions.
Although it has not been possible in the high multiplicity
environment of Au+Au collisions at RHIC to fully reconstruct jets, these high-$p_T$ particles
should be good surrogates for the jets. As shown in Fig.~\ref{fig:jet_suppression}, $\pi^0$'s
are suppressed by more than a factor of two in central Au+Au collisions, but have little or no
modification for the cold-nuclear matter control measurement in d+Au collisions. This is interpreted as evidence
for large energy losses of the jets in the final-state where they traverse the hot high-density matter
created in the Au+Au collisions. Models infer from this that matter densities of $\sim{15}$ times
normal nuclear matter density are involved in the earliest stages, before expansion, of the created medium. 

\begin{figure}
 \includegraphics[width=\linewidth]{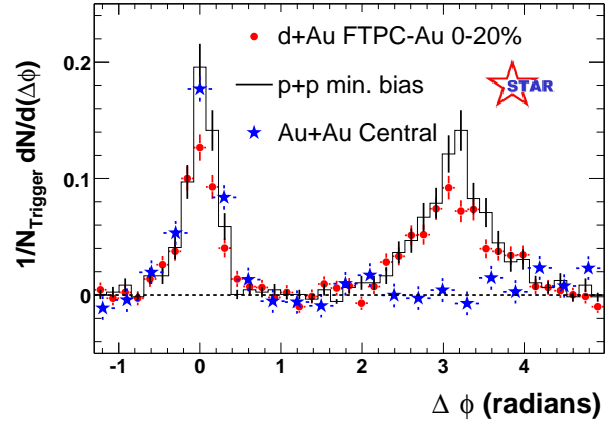}
\caption{Distribution in azimuthal angle ($\phi$), with-respect-to a tagging high-$p_T$
($p_T = 4 - 6$ GeV/c)
particle, of other particles ($p_T > 2$ GeV/c) seen in the same event for central Au+Au
collisions (blue stars), compared to d+Au collisions (red dots)
and p+p collisions (black histogram and vertical bars)\cite{star_jet_correlations}.}
\label{fig:jet_correlations}       
\end{figure}

In addition to studying the yield of hadrons vs $p_T$, the correlation of hadron pairs has also been
studied. In this case a ``tag" is provided by one high-$p_T$ hadron, and then the distribution of
other hadrons in the same event is observed. Fig.~\ref{fig:jet_correlations} shows these correlations
for central Au+Au, central d+Au and p+p collisions.
In d+Au collisions the correlation is essentially unaltered from that for p+p collisions,
while for central Au+Au collisions the ``away-side" correlation ($\Delta\phi \sim \pi$) disappears,
with the ``near-side" peak (other particles associated with the jet that the tagging
particle is from) remains like that of p+p and d+Au. This result is usually interpreted as
a tagging jet coming from near the surface and a partner ``away-side" jet that is degraded by the thick
high-density medium it has to pass through on the other side. Further studies looking at lower momenta for the
``away-side" hadrons appear to find remnants of this jet in broad distributions of low momentum particles.
In some momentum windows these ``away-side" particles even show a very interesting split
distribution with a depression in their yield for exactly back-to-back angles, and side maxima.
This phenomena has been interpreted by some as evidence for ``mach-cone" effects in the
medium\cite{mach_cone}.

\section{Direct Photons and the Initial State}
\label{sec:4}
\begin{figure}
  \includegraphics[width=\linewidth]{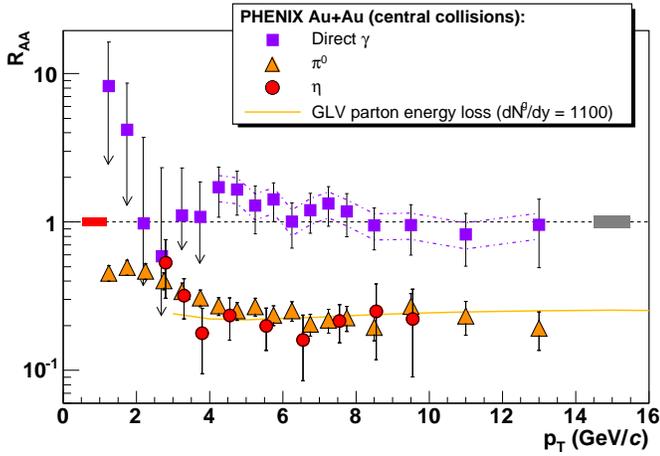}
\caption{Direct photons (purple squares) are not suppressed in central Au+Au collisions
as compared to the strong suppression seen for $\pi^0$'s (yellow triangles) and $\eta$'s (red circles)
\cite{phenix_photon}.The solid curve is a parton energy loss calculation\cite{vitev_dedx}.}
\label{fig:photon_vs_jet}       
\end{figure}

Direct photons, although difficult to isolate experimentally, because of their weak electromagnetic
interaction with the medium in the final state, connect directly to the
initial state where the hard interactions take place - before the thermalization of the
medium. In Fig.~\ref{fig:photon_vs_jet} the direct photons in central Au+Au collisions
show that the initial state is not modified from that of p+p collisions, while the neutral
mesons (as discussed above) are strongly suppressed due to final-state effects in the hot
dense medium. This comparison affirms the idea discussed above, that the observed hadron jet
modifications do come from the final state.

\section{Heavy Quarks}
\label{sec:5}
\begin{figure}
\begin{center}
  \includegraphics[height=80mm]{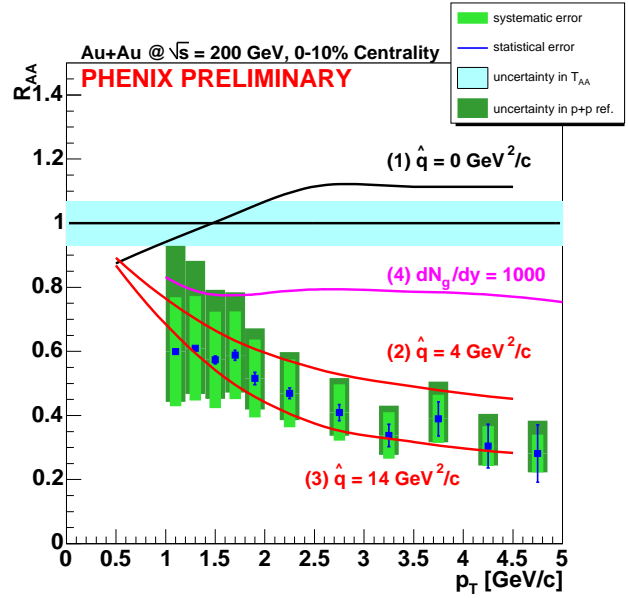}
\caption{Nuclear modification factor, $R_{AA}$, for heavy-quarks at central rapidity observed
using non-photonic electrons in central Au+Au collisions at 200 GeV\cite{phenix_raa_charm}.}
\label{fig:charm_central}       
\end{center}
\end{figure}

\begin{figure}
\begin{center}
  \includegraphics[height=60mm]{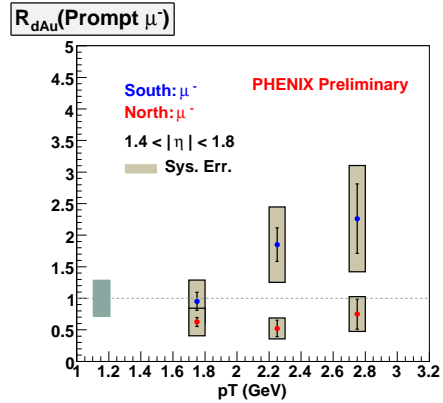}
\caption{Nuclear suppression factor, $R_{dAu}$, at forward and backward rapidity for heavy-quarks observed
using prompt muons (heavy quarks) in d+Au collisions at 200 GeV\cite{phenix_rdau_charm}.}
\label{fig:charm_forward}       
\end{center}
\end{figure}

So far we have discussed light hadrons, including the large energy loss observed
in the final state when they pass through the high-density matter created in central
Au+Au collisions. Heavy quarks are expected to suffer different effects in this medium,
but they are more difficult to measure both because of their smaller production cross sections 
and due to the difficulty of separating them from backgrounds. Despite these difficulties,
initial measurements primarily through measurements of leptons from the semi-leptonic decay
of the heavy (charm and beauty) mesons are now available. In Fig.~\ref{fig:charm_central}
the suppression of non-photonic electrons at mid-rapidity from charm and beauty decays is shown. The
large suppression seen, like that for light hadrons, is interpreted as energy loss in the high-density
medium created in these central Au+Au collisions. Similar measurements at forward rapidity are
being worked on using decays to muons, but so far results are only available for d+Au and p+p
collisions. The latter are shown in Fig.~\ref{fig:charm_forward} where one sees that d+Au
heavy-quark yields are suppressed at forward rapidity (small-x or shadowing region in the Au
nucleus) and enhanced at backward rapidity (larger-x or anti-shadowing region). The suppression
in the forward or small-x region is usually attributed to nuclear shadowing of the gluon
distributions\cite{shadowing} or to gluon saturation models\cite{CGC}; but could also involve other cold nuclear
matter effects such as initial-state gluon energy loss.

\begin{figure}
  \includegraphics[height=60mm]{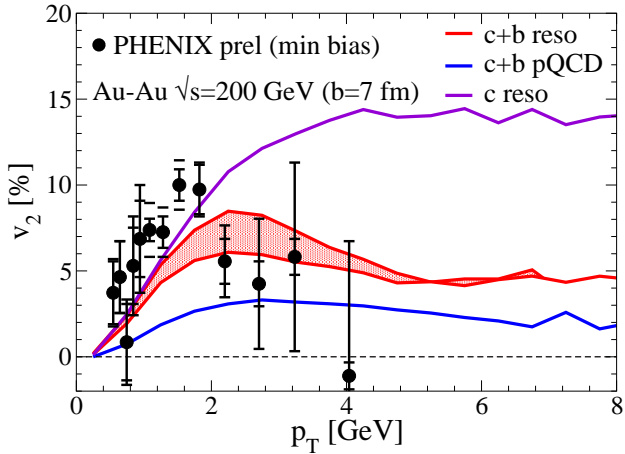}
\caption{Flow of heavy quarks at mid-rapidity for 200 GeV minimum-bias Au+Au collisions, as obtained
from non-photonic electrons\cite{phenix_raa_charm}. Data is compared to theoretical calculations\cite{rapp_flow} that
have no flow (blue curve), only charm flow (purple), or both charm and beauty flow (red).}
\label{fig:charm_flow}       
\end{figure}

The electrons from heavy-quark decays have also been analyzed in terms of their elliptic flow, analogous
to what was discussed above for hadrons. Surprisingly, at least for small $p_T$ values, these
heavy quarks also seem to exhibit flow (non-zero $v_2$), as shown in Fig.~\ref{fig:charm_flow}.
Although the data uncertainties remain large due to the low rates of heavy-quark production
and the large systematics background subtraction, the usual interpretation of
these results is that for the lowest transverse momentum the charm quarks do
flow with the light quarks, but as the momentum goes up they punch through the thermalized
medium and the flow disappears. However, more accurate measurements will be needed to
firm up the true characteristics of heavy-quark flow (Section \ref{sec:7}).

\section{$J/\psi$s and Color Screening in the Medium}
\label{sec:6}
The $J/\psi$ and other heavy quarkonia are thought to provide a key signature for a
deconfined medium. Early predictions were that the two heavy quarks that would form the
bound state would be screened from each other in the high-density deconfined medium\cite{matsui_satz}.
This effect would depend on the size or binding energy of the specific state
and so different states ($J/\Psi$, $\psi$', $\chi_C$, $\Upsilon_{1S}$, $\Upsilon_{2S}$,
$\Upsilon_{3S}$) would ``melt" at different energy densities. However, more recently,
lattice QCD calculations have suggested that in fact the $J/\psi$ would not be screened
unless effective densities of the hot plasma created in these heavy-ion collisions
exceeded twice the critical temperature\cite{lattice_jpsi}.

\begin{figure}
\begin{center}
  \includegraphics[height=140mm,bb=213 38 519 550,clip]{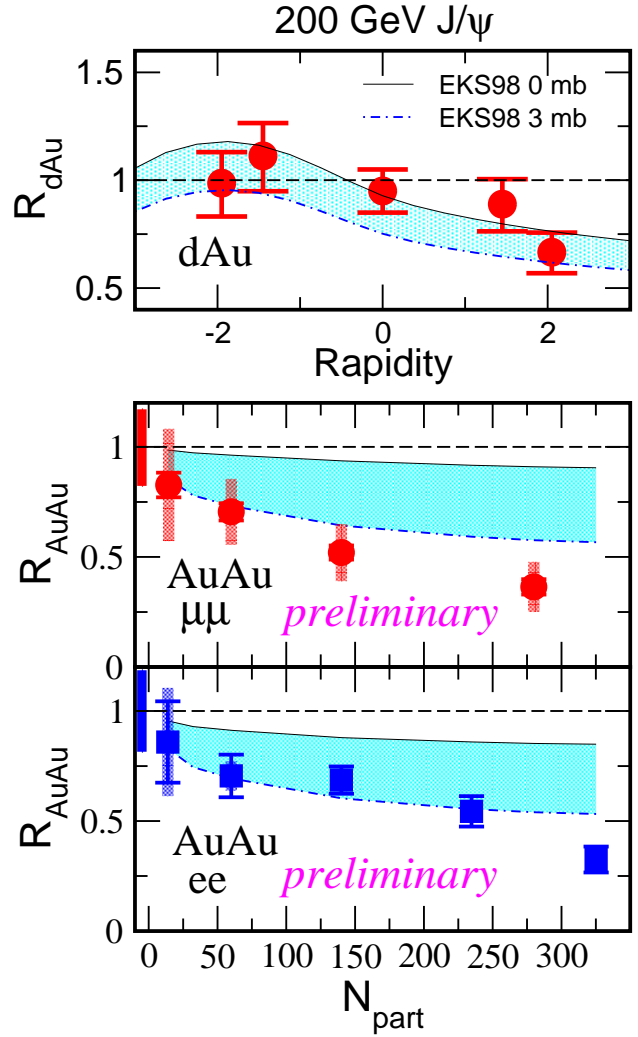}
\caption{Cold nuclear matter effects on $J/\Psi$ production are constrained by
the limited accuracy of our present d+Au data\cite{ppg038} (blue band in top panel). Corresponding
calculations for Au+Au collisions give bands on the bottom panels for forward rapidity(middle panel) and central
rapidity (bottom panel) data\cite{phenix_auau_jpsi}. Where $N_{part}$ is the number of participants,
and is a measure of the centrality of the Au+Au collision.}
\label{fig:rdau_aa_eks98_band}       
\end{center}
\end{figure}

The $J/\psi$ suppression at RHIC was predicted to be larger than that observed at the SPS by
most of the theoretical models that were successful in describing the SPS data\cite{jpsi_theories,rapp}.
This reflects
the expectation that the matter created at RHIC would be hotter and longer-lived
than that for the lower energy SPS measurements. Contrary to this expectation, the
measurements at RHIC show suppression very similar to that at lower energy.
In Fig.~\ref{fig:rdau_aa_eks98_band} are shown preliminary $J/\psi$ measurements for
Au+Au collisions from PHENIX\cite{phenix_auau_jpsi}, along with earlier measurements of the
cold nuclear matter effects as seen in d+Au collisions\cite{ppg038} at the same energy.
First, one can see that
simple models that describe the poor statistics d+Au measurements for a range of absorption
cross sections give a large uncertainty in the expected cold nuclear matter effects for
Au+Au collisions (blue bands). Clearly, better d+Au data is needed to more accurately constrain
the cold nuclear matter effects and to allow a more quantitative understanding of how much
of the Au+Au suppression does not come from these effects, and how much may come from
additional effects of the hot-dense matter. Nevertheless, for the most
central collisions, the Au+Au data clearly show a stronger suppression than one would expect
from cold nuclear matter effects alone.

\begin{figure}
  \includegraphics[height=60mm]{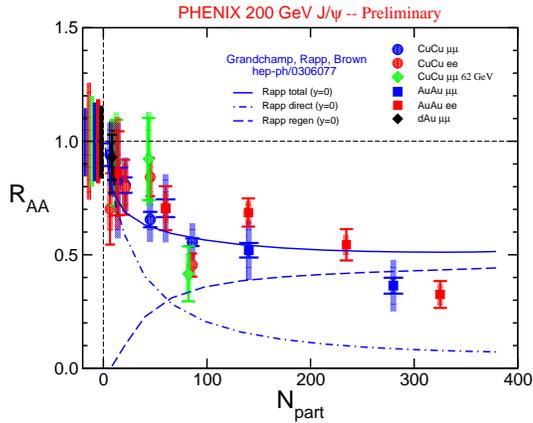}
\caption{Comparison vs centrality of the Au+Au $J/\psi$ data\cite{phenix_auau_jpsi} to theoretical calculations
that include both dissociation of the $c{\bar{c}}$
by a large gluon density and regeneration effects\cite{rapp}.}
\label{fig:raa_npart_rapp}       
\end{figure}

At present, we are left with two theoretical pictures which both provide a plausible explanation
of the level of suppression seen in the RHIC data. One of these, shown in Fig.~\ref{fig:raa_npart_rapp},
was actually a prediction from before the experimental results were obtained\cite{rapp}.
This model
includes strong dissociation of the charm pairs from the large gluon density created in the
collisions (analogous to screening), but also includes regeneration of bound charm pairs ($J/\psi$'s)
in the later stages of the expansion driven by the large production and high density of
independently produced charm quarks. So in this model, there is a stronger ``screening" effect at
RHIC than at the SPS but it is compensated for by the regeneration, resulting in a net suppression very similar
to that seen at the SPS.

\begin{figure}
\begin{center}
  \includegraphics[height=60mm]{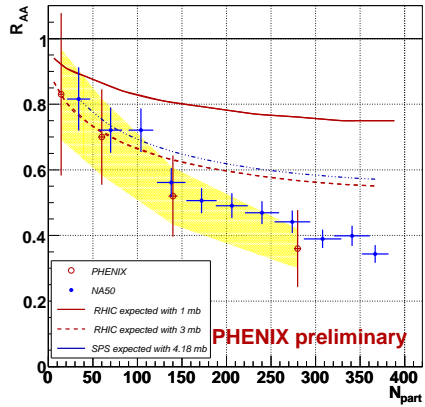}
\caption{200 GeV RHIC $J/\psi$ data\cite{phenix_auau_jpsi} and 17.4 GeV NA50\cite{na50_jpsi}
data plotted together vs $N_{part}$,
showing the universal dependence on number of participants. The curves
are calculations for RHIC with shadowing and absorption cross sections of
1 mb (solid red) and 3 mb (dashed red)\cite{vogt_RAA}; and for NA50 with an
absorption cross section of 4.18 mb (solid blue).}
\label{fig:moriond_na50_sps_2}       
\end{center}
\end{figure}

The second picture, sequential screening\cite{sequential_screening},
asserts that (as suggested by recent lattice calculations)
the $J/\psi$ is not screened in central A+A collisions at RHIC or at SPS energies. Rather
the observed suppression comes only from screening of the higher mass states, $\chi_C$ and
$\psi\prime$, that through feed down normally provide about $\sim 40\%$ of the $J/\psi$ production.
This picture provides a simple explanation for the nearly identical suppression observed at RHIC
and the SPS seen in Fig.~\ref{fig:moriond_na50_sps_2}. The $J/\psi$ itself would be screened
only for higher energy collisions at the Large Hadron Collider (LHC).

\section{Future - RHIC-II and Detector Upgrades}
\label{sec:7}
The luminosities obtained at RHIC are just beginning to reach levels where the more rare
probes such as the $\Upsilon$ can be studied. In the future
this will be improved further when RHIC-II, with its electron cooling, allows even higher
luminosities. In addition upgrades of the RHIC detectors are underway, most notably adding
high resolution silicon vertex tracking in the inner regions of the large detectors (PHENIX
and STAR) in order to make identification of heavy quarks more explicit via the separation of the
primary and secondary vertices associated with the creation, propagation and subsequent decay
of the heavy mesons. These detached vertex measurements of heavy quarks, along with the increasing
luminosity will dramatically reduce systematic and statistical uncertainties in these measurements
and will enable exclusive measurements of heavy quarks such as $B \rightarrow J/\psi X$.

\section{Summary}
\label{sec:8}
In summary, evidence is mounting at RHIC for the creation of a new form of matter that is
1) dense and gives large energy losses and modifications of jet correlations for hadrons,
2) appears to be thermalized very early and to exhibit maximal flow as predicted by hydrodynamics
models, with quark degrees of freedom, 3) also causes large energy loss and flow for heavy
quarks, and 4) causes strong suppression, beyond that expected from cold nuclear matter effects,
for the $J/\psi$. Future luminosity and detector upgrades will enable firming up and extending
our understanding of these phenomena.

%
%

\end{document}